\documentclass[useAMS]{mn2e}
\usepackage{times,psfig} 
 
\begin{document}
\title[Sedimentation of helium]{Effects of sedimented helium on the
X--ray properties of galaxy clusters}


\author[]
{\parbox[]{6.in} {S. Ettori$^1$ and A.C. Fabian$^2$\\
\footnotesize
$^1$ INAF, Osservatorio Astronomico di Bologna, via Ranzani
  1, I-40127 Bologna, Italy (stefano.ettori@oabo.inaf.it) \\
$^2$ Institute of Astronomy, Madingley Road, Cambridge CB3 0HA \\
}}                                            
\date{Accepted 2006 March 13.  Received 2006 March 13; in original form 2005 December 6}
\maketitle

\begin{abstract}
In this Letter, we consider the role played by the sedimentation 
of helium nuclei on the emissivity and metallicity distribution 
of the X-ray emitting plasma in the cores of relaxed galaxy clusters.
We model the gas density and temperature profiles of 
nearby cooling core clusters to estimate the gravitational
acceleration acting on the helium and show that its sedimentation
time scale is too long with respect to the present age of these
objects to play a significant role.
However, we argue that these time scales have to be definitely
lower in the past allowing the helium to settle in the cluster
cooling cores and raise its abundance to values higher than 
the solar one. A direct consequence of this speculation is that
the helium, by increasing the total X-ray emissivity,
reduces the measured metal abundance in the inner ($r \la 20$ kpc) 
cluster regions.
\end{abstract}


\section{INTRODUCTION} 

The plasma in X-ray emitting clusters of galaxies is composed
by hydrogen (H), helium (He) and various ionized metals of which
iron (Fe) is the most interesting from the astrophysical point of view
owing to its connection to the star formation activity 
(e.g. Renzini 2003).
In an X-ray emitting plasma with solar abundance, 
for each atom of hydrogen a number density of $9.77 \times 10^{-2}$ 
ions of He and $4.68 \times 10^{-5}$ ions of Fe is expected
(as in Anders \& Grevesse 1989; for the most recent determinations
in Grevesse \& Sauval 1998 and Asplund, Grevesse \& Sauval 2005, 
$8.51 \times 10^{-2}$ ions of He correspond to each atom of H,
whereas the number density of Fe is  
$3.16 \times 10^{-5}$ and $2.82 \times 10^{-5}$, respectively).
Being the atomic weight $A$ of (H, He, Fe) $= (1, 4, 56)$,
the mass fraction on the total is $0.75$ for the hydrogen
(generally labeled as $X$), $0.24$ for helium (called $Y$)
and the rest ($Z = 1-X-Y \approx 0.017$) for heavier elements.
 
\begin{figure*}
\hbox{
  \psfig{figure=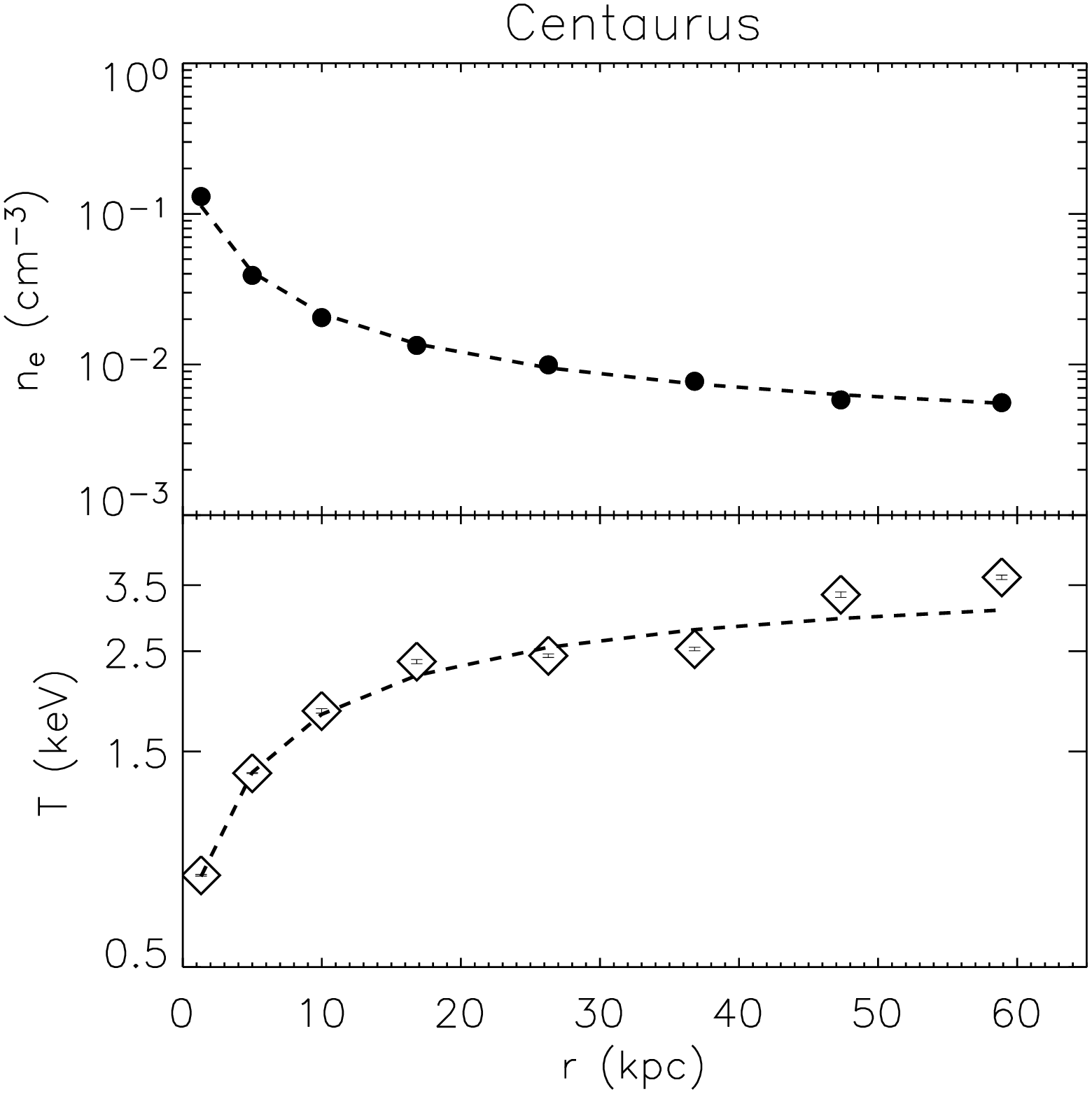,width=.33\textwidth}
  \psfig{figure=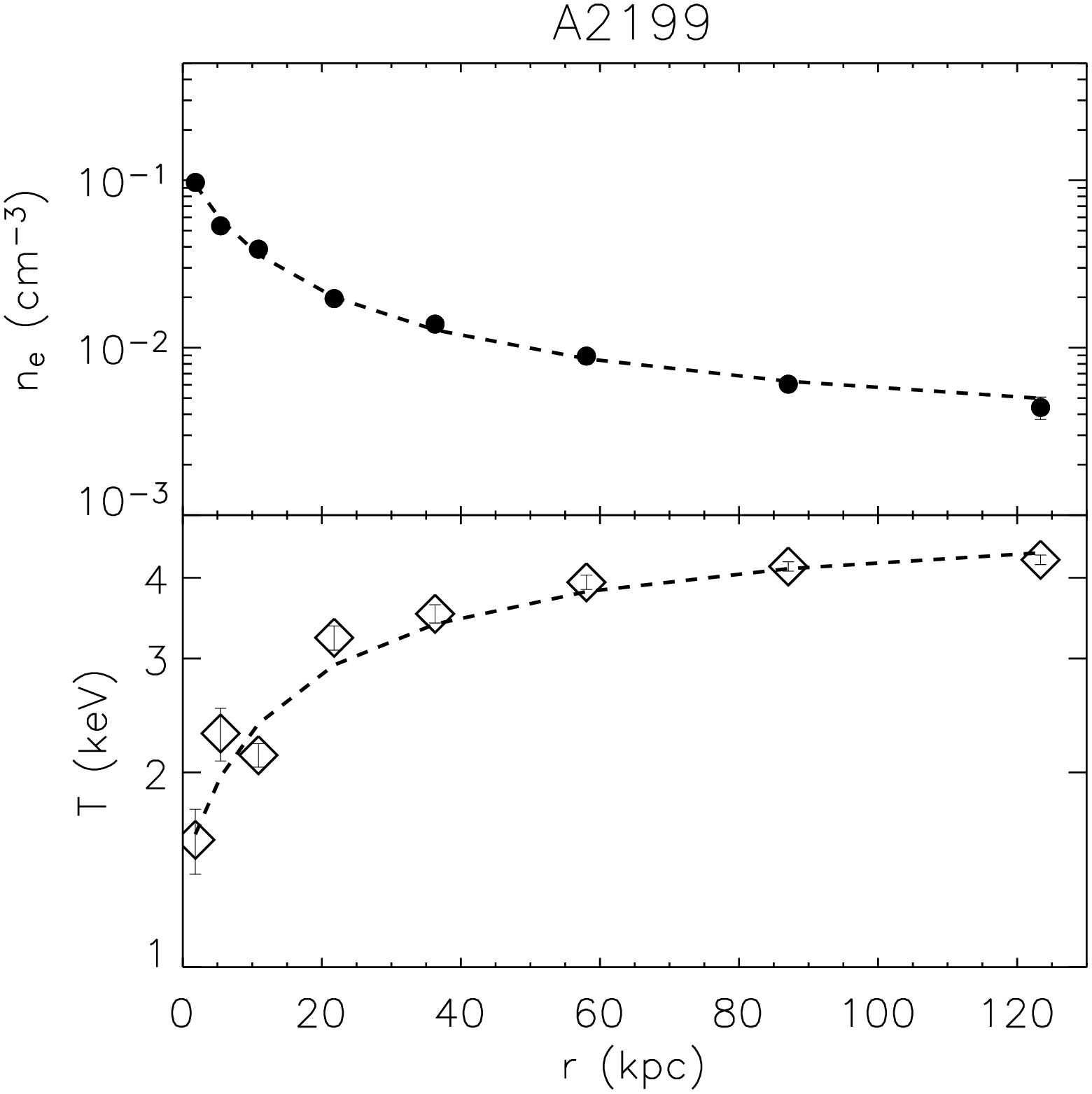,width=.33\textwidth}
  \psfig{figure=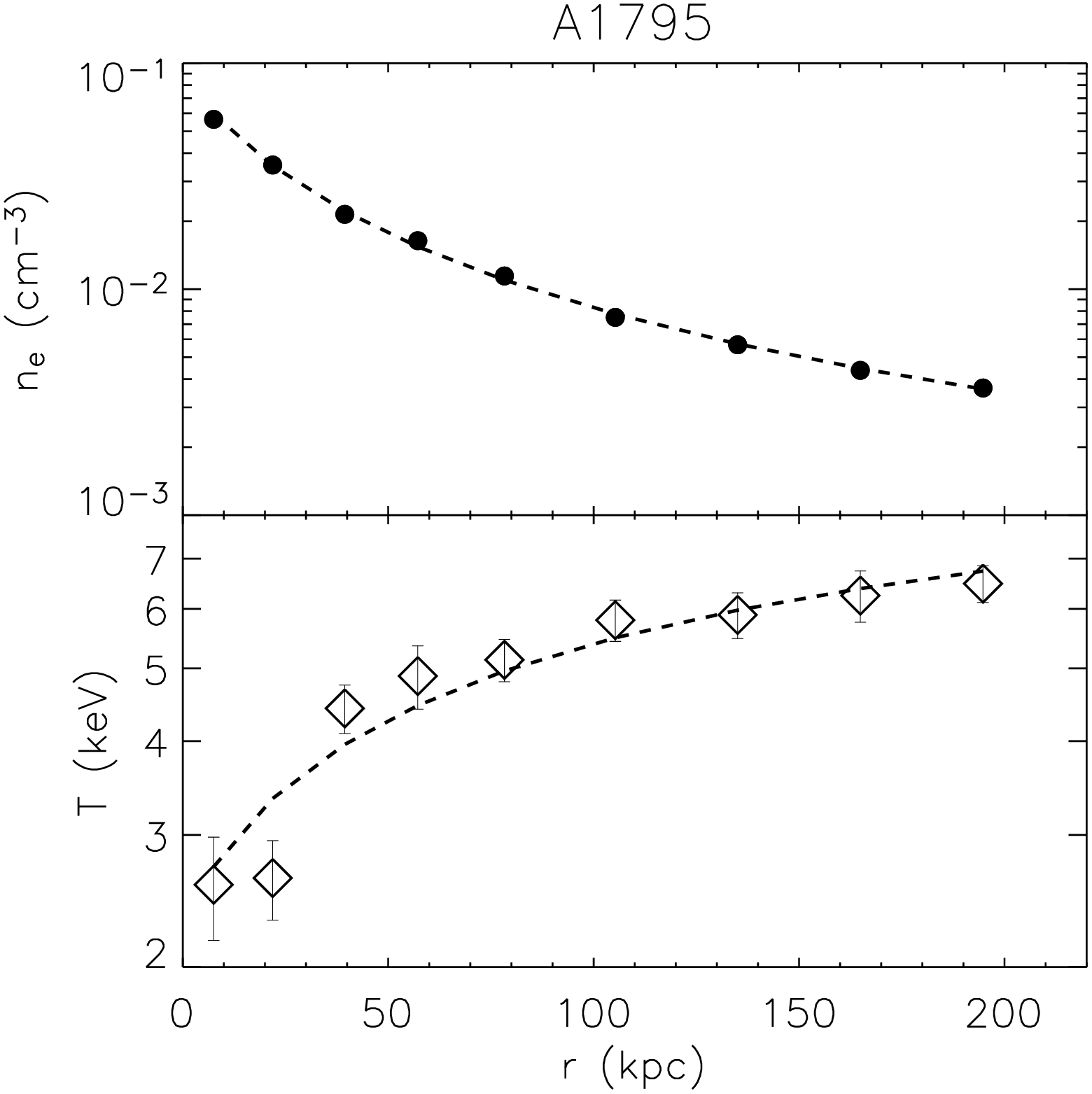,width=.33\textwidth}
}
\caption{Joint best-fit of the gas density and temperature profile
with a modified NFW gas profile of the deprojected data of Centaurus
({\it left}, Sanders et al. 2002), A2199, ({\it right}, Johnstone et al. 2002)
and A1795 ({\it center}, Ettori et al. 2002). 
The density profile presents a mean relative uncertainty of about 
2 per cent that is 5 times lower than the value associated
with the estimates of the temperature profile.
} \label{fig:nt} \end{figure*}

Diffusion of helium and other metals can occur in the core
regions of the intracluster plasma under the attractive action 
of the central gravitational potential, enhancing their 
abundances on time scales comparable to the cluster age.
Qin \& Wu (2000) and Chuzhoy \& Nusser (2003), 
following the work by Fabian \& Pringle (1977), 
Rephaeli (1978), Abramopoulos et al. (1981), Gilfanov \& Syunyaev (1984) on 
the sedimentation of elements in clusters, studied the concentration
of helium in cores concluding that it can gravitationally settle 
within a Hubble time.
Abramopoulos et al. (1981) obtained the same results of Qin \& Wu
assuming a global thermal equilibrium and the Boltzmann equation, while
Gilfanov \& Syunyaev (1984) applied the diffusion equation to the light 
elements (atomic number $Z_i < 16$) where the changes are more efficient.
Therefore, in a H-He plasma, the helium tends to be more centrally peaked
than the hydrogen.

\begin{figure*}
\hbox{
 \psfig{figure=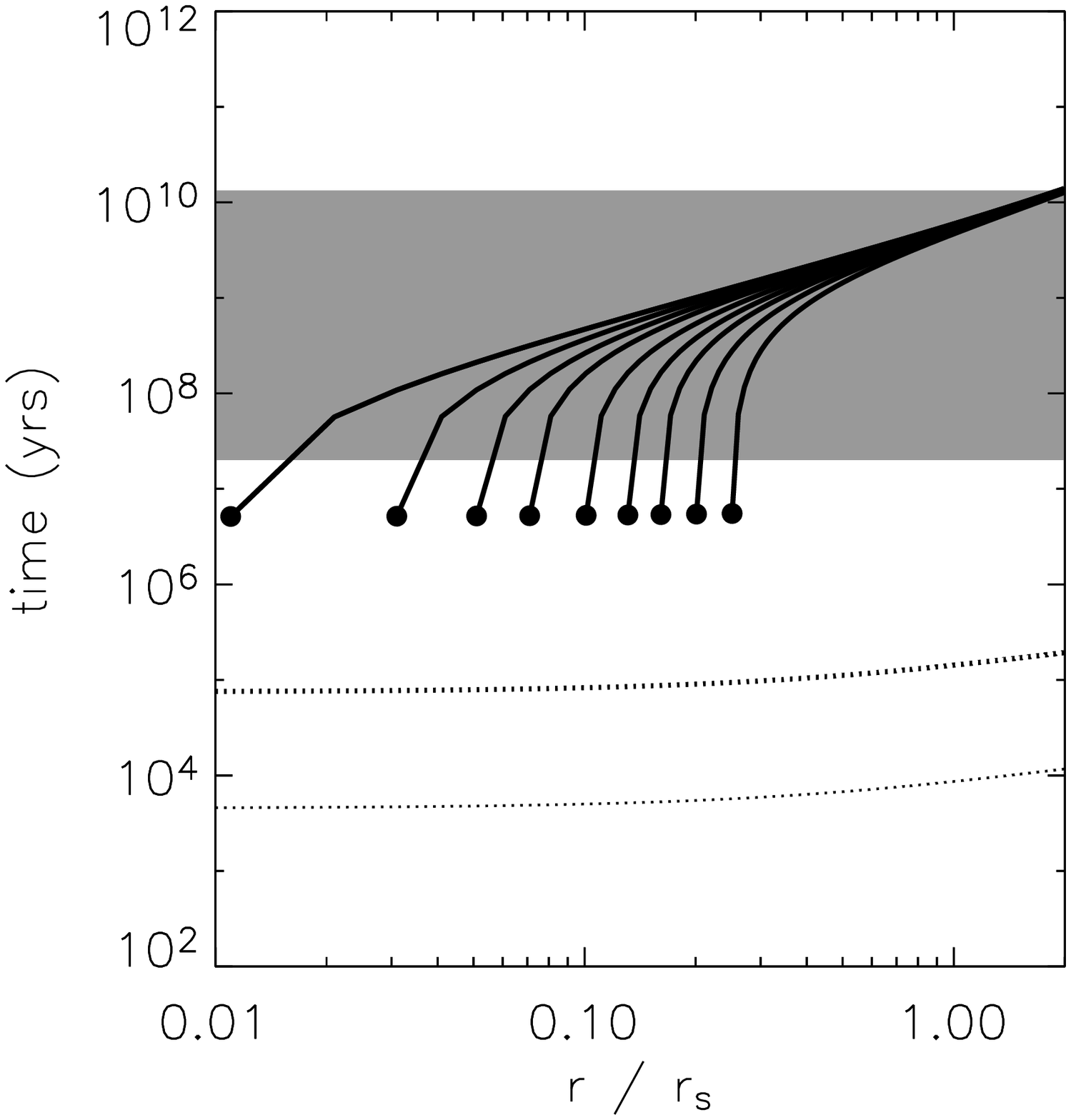,width=.45\textwidth}
 \psfig{figure=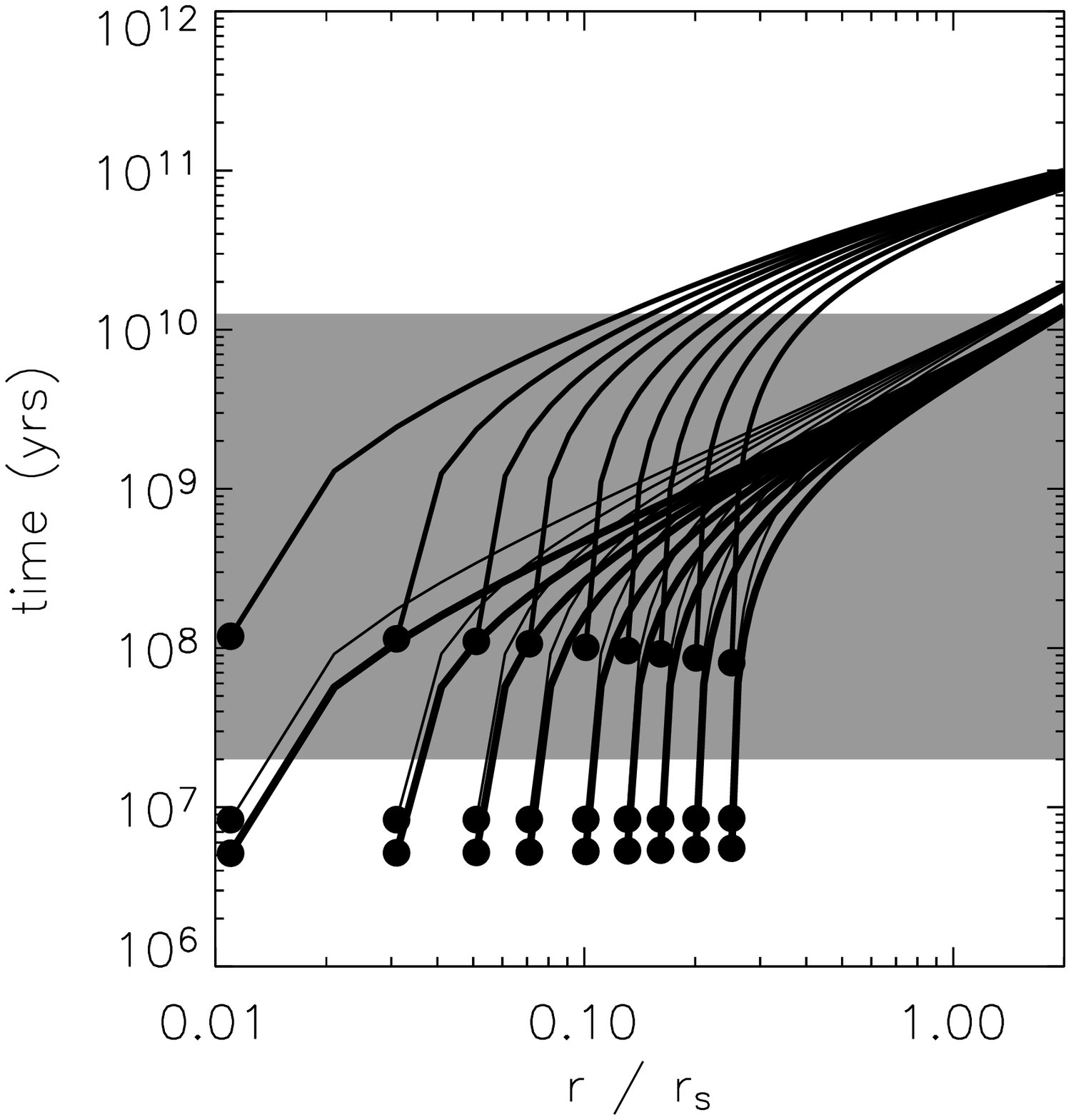,width=.45\textwidth}
}
\caption{{\it (Left panel)} 
Equipartition time between H and He (thick dotted line) and
between H and Fe (thin dotted line) in the Centaurus cluster; 
sedimentation times, $t_{\rm sed} = \int^{r_{\rm out}}_{r_{\rm in}}
dr / v_{\rm sed}(r)$, of He (solid line).
The dots mark the radius $r_{\rm in}$ at which the material is accumulated from
the regions beyond with a sedimentation time represented by the line.
{\it (Right panel)}
Sedimentation times of He for the three objects under examination 
(from thickest to thinnest line: Centaurus, A2199, A1795).
The cluster potentials are described by equation~\ref{eq:pot}
given the best-fit values in Table~1. 
A flat temperature profile is assumed (see text at the end of Section~2).
The shaded region ranges between $2 \times 10^7$ years, which is about the cooling 
time associated with a gas with solar abundance of helium and proton density of 0.2
cm$^{-3}$ at $\sim 1$ keV. 
and the age of the Universe for the assumed cosmology, $1.3 \times 10^{10}$ years. 
} \label{tv_sed} \end{figure*}

More recently, Chuzhoy \& Loeb (2004) solved the full diffusion
equations derived by Burgers (1969) for a multicomponent
fluid, finding that the diffusion, if the suppression due to 
magnetic fields is modest, can alter X-ray properties as the
steepening inward of the baryon distribution as well as the 
spectrum and evolution of stars forming out from helium-rich gas.
However, all these studies assume a gas temperature constant in time and space
in evaluating the gravitational acceleration in the cluster core,
where instead steep spatial gradients are observed in cool core
clusters.

In the present work, we analyze the effects of sedimented 
elements, He in particular, on the description of the X-ray 
emitting plasma that is generally assumed with a fixed solar
abundance of helium (as tabulated in, e.g., Anders \& Grevesse 1989; 
on more recent estimates of the solar chemical composition see, for example,
Asplund, Grevesse \& Sauval 2005) and a relative contribution from metals 
determined through the equivalent width of the detected
emission lines in X-ray spectra.
We model properly the central steepening of the observed gas
temperature and density profiles to estimate the gravitation potential
in the inner cluster regions where the effect of the gravitational
sedimentation is expected to be more relevant.
In these regions, we prove that the underestimation of the amount of helium
can propagate to inaccurate measurements of the total X-ray emissivity,
gas density and metallicity.

\section{Sedimentation of helium}

Spitzer (1956) presents the physics of fully ionized gases
that we adopt in the following discussion.
In general, the time of equipartition between H and He (Fe) ions is low
when it is compared to the cluster age, whereas the sedimentation time
is comparable to it. Hereafter, we assume a cosmology $(H_0, 
\Omega_{\rm m}, \Omega_{\Lambda}) = (70 \ {\rm km} \ {\rm s}^{-1} 
{\rm Mpc}^{-1}, 0.3, 0.7)$ that implies an age of the Universe
of $1.3 \times 10^{10}$ years.

To describe the action of the gravitational acceleration 
in drifting inward the plasma elements, 
we consider an intracluster gas in hydrostatic equilibrium 
with a Navarro, Frenk \& White (1995, hereafter NFW) potential 
\begin{equation}
\frac{d \phi}{dr} = 4 \pi G \rho_{\rm s} r_{\rm s} 
\left( \frac{\ln(1+x)}{x^2} - \frac{1}{x (1+x)} \right) 
= 4 \pi G \rho_{\rm s} r_{\rm s} \; f(x),
\label{eq:nfw}
\end{equation} 
with $x = r / r_{\rm s}$.
We are interested to measure such potential in the inner regions
of a cluster, where a steep gradient in temperature is generally
observed. 
A model for the gas density profile obtained analytically
from the hydrostatic equilibrium equation (HEE)
with a NFW potential is available in literature
(e.g. Makino, Sasaki \& Suto 1998, Ettori \& Fabian 1999), 
but it assumes an isothermal gas. We relax this assumption 
to accommodate the observed gradients and modify 
accordingly the model of the gas density profile. 
We assume a temperature profile with the functional form
\begin{equation}
T_{\rm gas} = T_{\rm max} \frac{x^b + T_0/T_{\rm max}}{x^b + 1}
 = T_{\rm max} \; t(x)
\label{eq:tgas}
\end{equation}
that is just the rearranged expression of the equation that
Allen, Schmidt \& Fabian (2001) show to be able to reproduce the
temperature profiles of relaxed clusters (see Vikhlinin et al. 2005).
We use then this model in the HEE with a NFW potential
to recover a numerical expression of the gas density profile:
\begin{equation}
n_{\rm gas}(x) = n_{\rm gas}(0) \frac{1}{t(x)} 
\exp \left(-\eta \int_0^x \frac{f(x)}{t(x)} dx \right),
\label{eq:ngas}
\end{equation} 
with $\eta = 4 \pi G \rho_{\rm s} 
r_{\rm s}^2 \mu m_{\rm p} / T_{\rm max}$.
 
These functional forms of the temperature and density 
profiles are then joint-fitted with a $\chi^2$ minimization
to the deprojected data of Centaurus (Sanders et al. 2002),
A2199 (Johnstone et al. 2002) and A1795 (Ettori et al. 2002).
The best-fit results are presented in Table~1 and 
plotted in Fig.\ref{fig:nt}. 

A proper fit of the gas density and temperature gradient 
allows us to define accurately the cluster gravitational 
acceleration $g$ by using the NFW potential in equation~\ref{eq:nfw},
\begin{equation}
g(r) = \frac{d \phi}{dr} = \frac{\eta T_{\rm max}}{\mu m_{\rm p} r_{\rm s}} f(r), 
\label{eq:pot}
\end{equation}
with the parameters $\eta$, $T_{\rm max}$
and $r_{\rm s} = r / x$ obtained as best-fit results
of the deprojected gas density and temperature profiles.
By using, instead, the best-fit parameters from an
isothermal NFW gas density profile, as generally done 
for this kind of analysis, one would infer a
potential in the inner regions that is larger than 
the adopted value by a factor of $\sim4$, mainly owing
to the assumed higher temperature value that propagates 
linearly to the gravitational acceleration.

For a Boltzmann distribution of particles labeled $1$ with 
density $n_1$ and thermal velocity $v_{\rm th} = 
(2 kT / A_1 m_{\rm p})^{1/2}$ in a plasma with temperature $kT$ in 
hydrostatic equilibrium with a NFW potential $g(r)$, the drift velocity 
of the heavier ions $2$ with respect to $1$ is given by
(e.g. Spitzer 1956)
\begin{equation}
v_{\rm sed}(r) = \frac{3 m_{\rm p}^2 \ A_1 A_2 \ v_{\rm th}^3 \ g(r)}
{16 \pi^{1/2} e^4 \ Z_1^2 Z_2^2 \ n_1 \ \ln \Lambda},
\end{equation}
where $\log \Lambda$ is the Coulomb logarithm equals 
to $37.9 + \ln \left( (kT / {\rm 10 keV})
(n_{\rm e, 1}/10^{-3} {\rm cm}^{-3})^{-1/2} \right)$.
Typical velocities of about $2$ kpc Gyr$^{-1}$ are expected which can make
reasonable the sedimentation of helium in the inner 10 kpc or so. 
The time elapsed in drifting from $r_{\rm inn}$ and $r_{\rm out}$
is the sedimentation time, $t_{\rm sed} = \int^{r_{\rm out}}_{r_{\rm in}}
dr / v_{\rm sed}(r)$, that is therefore proportional to
$n_1 \ g^{-1} \ kT^{-3/2}$.

\begin{table}
\caption{Best-fit results of the gas density and temperature profile
models as plotted in Fig.~\ref{fig:nt}}
\begin{tabular}{l c c c} 
\hline \\ 
 & Centaurus & A2199 & A1795 \\
\\
$z$  &  0.0104 & 0.0309 & 0.0632  \\
$r_{\rm s}$ (kpc) & 16 & 33 & 246  \\
$n_{\rm e}(0)$ (cm$^{-3}$) & 0.264 & 0.133 & 0.089  \\
$T_0$ (keV) &  0.48  &  1.36 & 2.25  \\
$T_{\rm max}$ (keV)  &  3.92  &  5.22 & 12.24  \\
$b$  &  0.91  &  0.94 & 0.86  \\
$\eta$ &  12.1  &  7.6 & 15.9  \\
\\
$\chi^2 (n_{\rm e})$ & 212 & 12 & 29 \\
$\chi^2 (T)$         & 453 & 20 & 10 \\
$\chi^2$/dof & 665/10 & 32/10 & 39/12 \\
\hline 
\end{tabular}
\end{table}

The time required to sediment helium in the inner cluster regions
is in the order of few Giga-years (see also Chuzhoy \& Loeb 2004).
Owing to the competing drag forces from the protons moving upwards and
the helium diffusing inwards, heavier elements can sediment with similar 
velocity, as discussed in Chuzoy \& Nasser (2003).
We conclude that sedimentation of helium is not an efficient
process, given the present condition of the cluster plasma. 

However, the intracluster medium evolved to a well-defined cool core
during its formation history. We can assume that the gas started originally
with a flat temperature distribution equal to the value measured
in the outskirts and a gas density profile obtained from the
equation~\ref{eq:ngas} by fixing $t(x)=1$ and by requiring that the
outer pressure remains unchanged as the overall gravitational potential.
While the evolution of a cool core from an initially isothermal plasma is
the natural consequence of the radiative processes taking place in the central
regions at higher density, there is not yet detailed studies from, e.g., 
cosmological simulations of structure formation that discuss the evolution
of cool cores owing to the still unknown balance of energetic losses and 
feedbacks that rules the X-ray emitting plasma in the cluster cores
(for example, the temperature profiles are expected
to be flatter and density higher at $z=1$ with respect to the local measurements
in Ettori et al. 2004, but Burns et al. 2004 suggest an alternative scenario
in which cool cores in massive clusters are formed through the merging of 
subclumps with its own cool core).
These assumptions imply that the gas temperature increases in the inner regions
by a factor of $\sim3$ and the density decreases in the center by about
an order of magnitude.
As shown in Fig.~\ref{tv_sed}, these changes in gas temperature and density
values cause the sedimentation time to be shorter. 
This makes reasonable the scenario
in which (i) the helium sedimented during the initial collapse and 
subsequent relaxation of the cluster plasma and 
(ii) it resides nowadays in super-solar abundance 
in cluster cores.


\begin{figure*}
\hbox{
  \psfig{figure=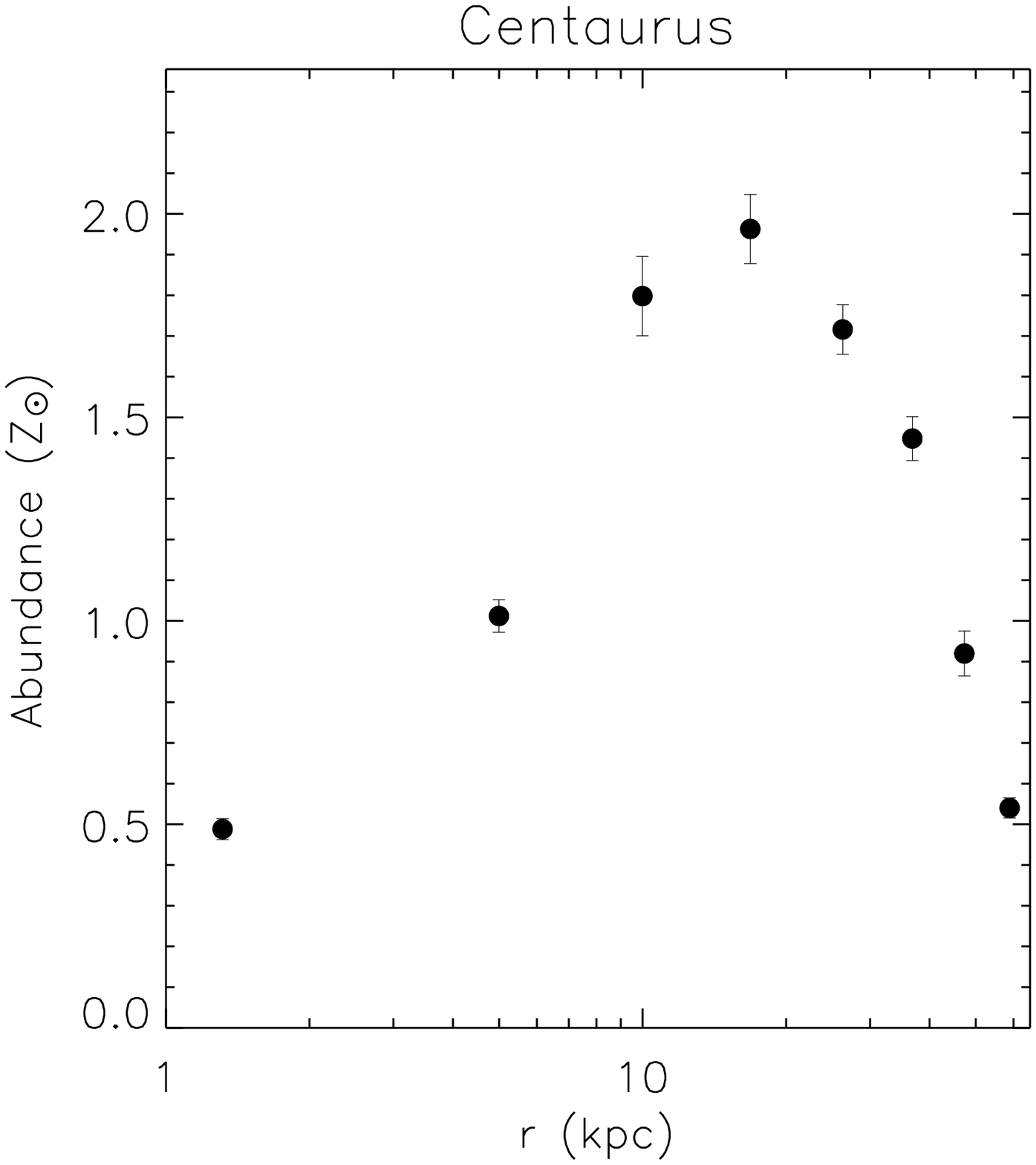,width=.33\textwidth}
  \psfig{figure=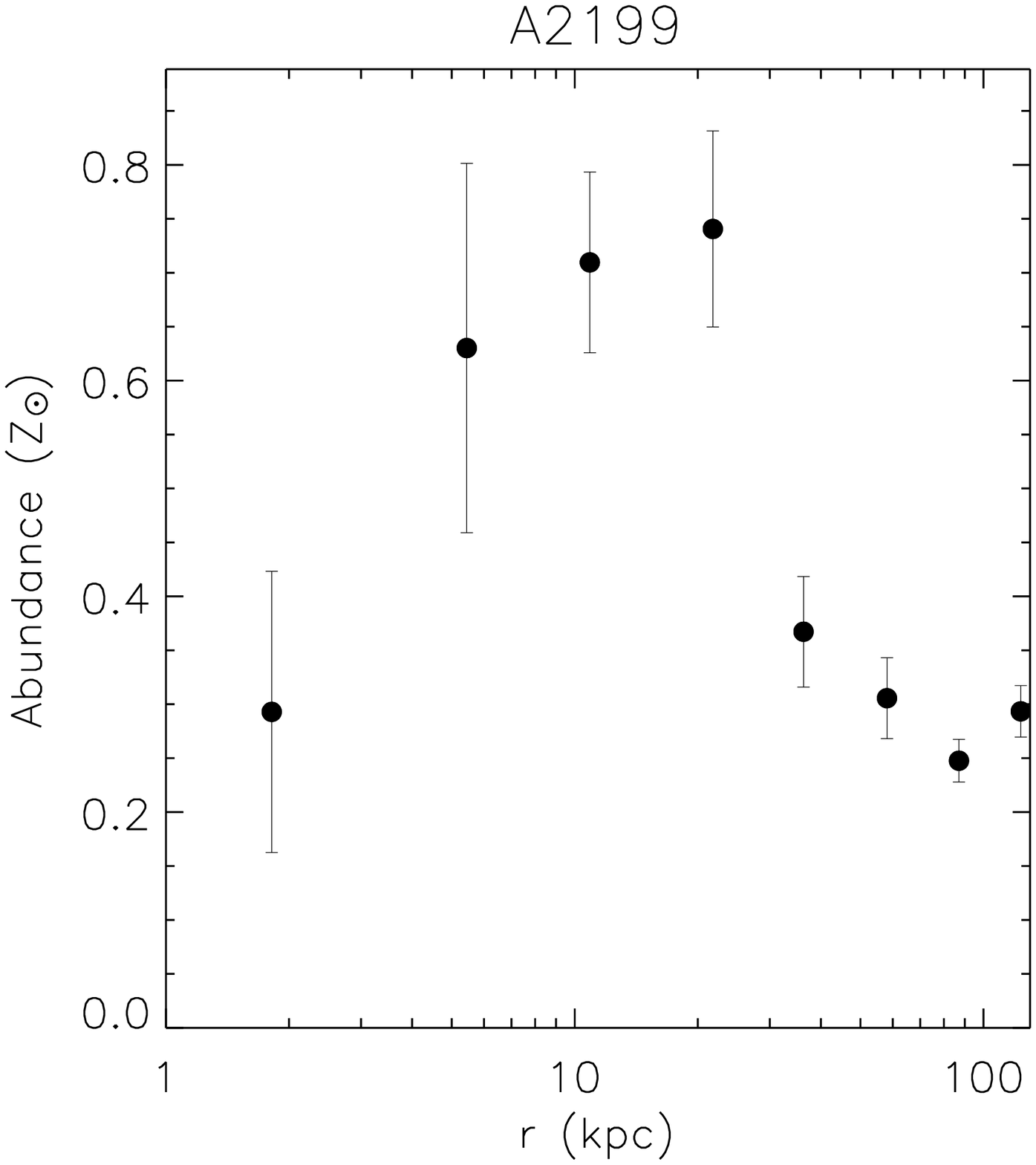,width=.33\textwidth}
  \psfig{figure=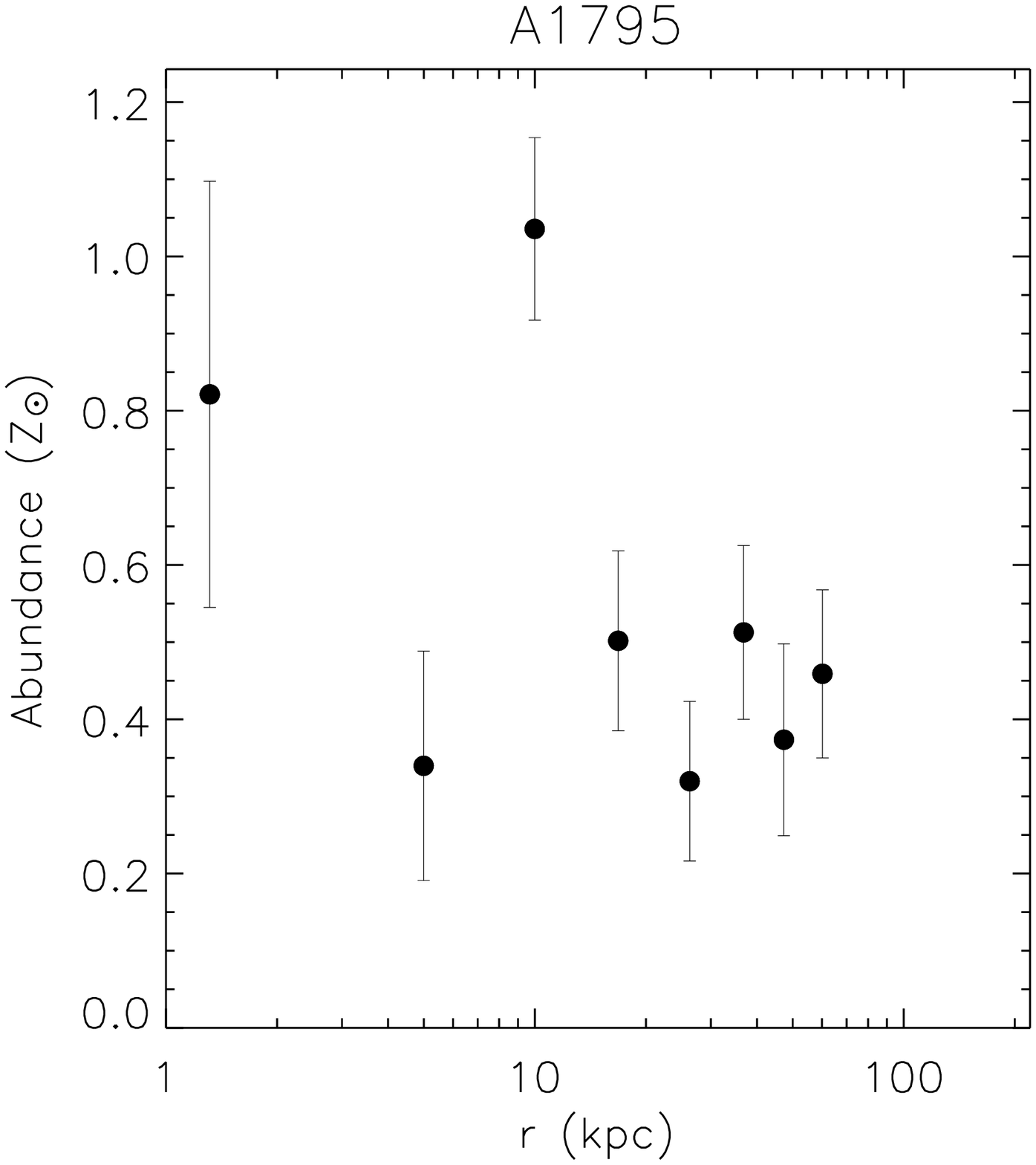,width=.33\textwidth} 
}
\caption{Deprojected metallicity profiles.
An inverted gradient is present in the inner cluster regions.
} \label{fig:abun} \end{figure*}

\section{Effects of the sedimented helium}

Intracluster plasma emits mainly by X-rays through bremsstrahlung 
with a typical emissivity, 
\begin{equation}
\epsilon \propto \Lambda(T_{\rm gas}) n_{\rm e} \sum M_i Z_i^2 n_i,
\label{eq:em}
\end{equation}
where $\Lambda(T_{\rm gas})$ is the cooling function of a plasma
with temperature $T_{\rm gas}$, $Z_i$ is the electric charge
(f.e. 1 for H, 2 for He, 24 for Fe), $M_i$ is the 
element abundance relative to the solar value and
$n_i$ is the density of the species in exam.
It is thus evident that any metal enhancement of its
relative abundance with respect to the hydrogen, as
occurs when sedimentation is relevant, 
affect the total X-ray emission.
It is important to note that the X-ray emissivity is generally
measured under the assumption that the helium abundance
is fixed to the solar value relatively to hydrogen,
i.e. $\epsilon \propto 1.85 n_{\rm p}^2$.
In the extreme case of a plasma composed by pure helium ($X=0, Y=1$),
$n_{\rm e} = 2 n_{\rm He}$ and $\epsilon \propto 8 n_{\rm He}^2$,
with the direct implication that if this plasma is assumed
with solar abundance, the total density is overestimated by a factor of 2. 
Moreover, the change in the relative abundance of H and He
induces variation in the number of electrons and ions present
in the plasma
\begin{equation}
\frac{n_{\rm e}}{n_{\rm p}} = c_M = \sum M_i Z_i N_i =
 1 +2 M_{\rm He} N_{\rm He} +\sum_{i \neq {\rm H, He}} M_i Z_i N_i,
\end{equation}
where $M_i$ is the element metallicity and $N_i$ is the particle
number density according to Anders \& Grevesse (1989),
and affects the measurements of the mean molecular weight,
\begin{equation}
\mu = \left(  \sum_i \frac{X_i (1+Z_i)}{A_i} \right)^{-1}
\approx \frac{1}{ 2X +(3/4)Y +(1/2)Z },
\end{equation}
that is generally fixed to a value of $\sim 0.6$ 
appropriate for solar abundance.
While $c_M$ and $\mu$ are quite independent of the metal
component of the plasma (e.g. $\mu = 0.618$ and $c_{M_i}=1.209$ 
for a solar abundance, $\mu = 0.613$ and $c_M=1.200$ for a 
metallicity of $0.3 Z_{\odot}$ for the reference values in 
Anders and Grevesse; note that, on the basis of the compilation
in Grevesse \& Sauval, $\mu = 0.605$ and $0.600$ and 
$c_M=1.182$ and $1.174$, for a solar and $0.3$ times solar 
abundance, respectively),
their values are highly sensitive to the variation in the helium
abundance: for $M_{\rm He}$ of, e.g., 2 and 10, 
$\mu = 0.699$ and $1$, respectively, $c_M=1.405$ and $2.968$.


Through these quantities, several other cluster measured 
properties are affected, like, e.g., the cooling time 
$t_{\rm cool} \approx n_{\rm g} T_{\rm g} / \epsilon \approx
(1+1/c_M) T_{\rm gas} / (n_{\rm p} \Lambda_{\rm cool})$,
the gas mass density $\rho_{\rm g}
= \mu (n_{\rm e} + n_{\rm p}) m_{\rm p}= 
\mu (1+c_M) n_{\rm p} m_{\rm p}$, 
the total gravitational mass measured through the
hydrostatic equilibrium equation, $M_{\rm tot} \propto \mu^{-1}$,
the gas mass fraction, $f_{\rm gas} = M_{\rm gas}/M_{\rm tot}
\propto \mu^2 (1+c_M)$.
The gas fraction, in particular, increases by a factor from 1.4 
to 4.8 for 2 and 10 times the solar abundance of helium, respectively.
To quantify the effect of sedimented helium on, e.g., the cooling time
we have first used the model {\tt vmekal} in XSPEC (Arnaud 1996)
to estimate the cooling function 
$\Lambda_{\rm cool} = \epsilon / (n_{\rm e} n_{\rm p})$
for a range of gas temperatures, $kT_{\rm gas}$, and He abundances, 
$M_{\rm He}$, in the interval 1 and 10 with respect to the solar chemical
compilation in Anders \& Grevesse (1989; see command {\tt abund} in XSPEC:
$n_{\rm He} = 9.77 \times 10^{-2} n_{\rm H}$).
As expected, higher helium abundance induces larger emission and shorter 
cooling time of the emitting plasma, with a cooling function that raises
for, e.g., a gas at 2 keV and proton density of 0.2 cm$^{-3}$
by a factor of 1.2 ($M_{\rm He}=2$) and 2.6 ($M_{\rm He}=10$) and 
a cooling time that decreases from $7 \times 10^7$ years
to $5.6 \times 10^7$ ($M_{\rm He}=2$) and $2 \times 10^7$ 
($M_{\rm He}=10$) years.
Under the assumption that bremsstrahlung dominates,
$t_{\rm sed}/t_{\rm cool} \propto n_{\rm gas}^2 / T_{\rm gas}^2$
which makes evident how the formation of a cool core, with 
the central gas that becomes progressively higher in density
and lower in temperature, reduces significantly 
any process of sedimentation.



\begin{figure*}
\hbox{
\psfig{figure=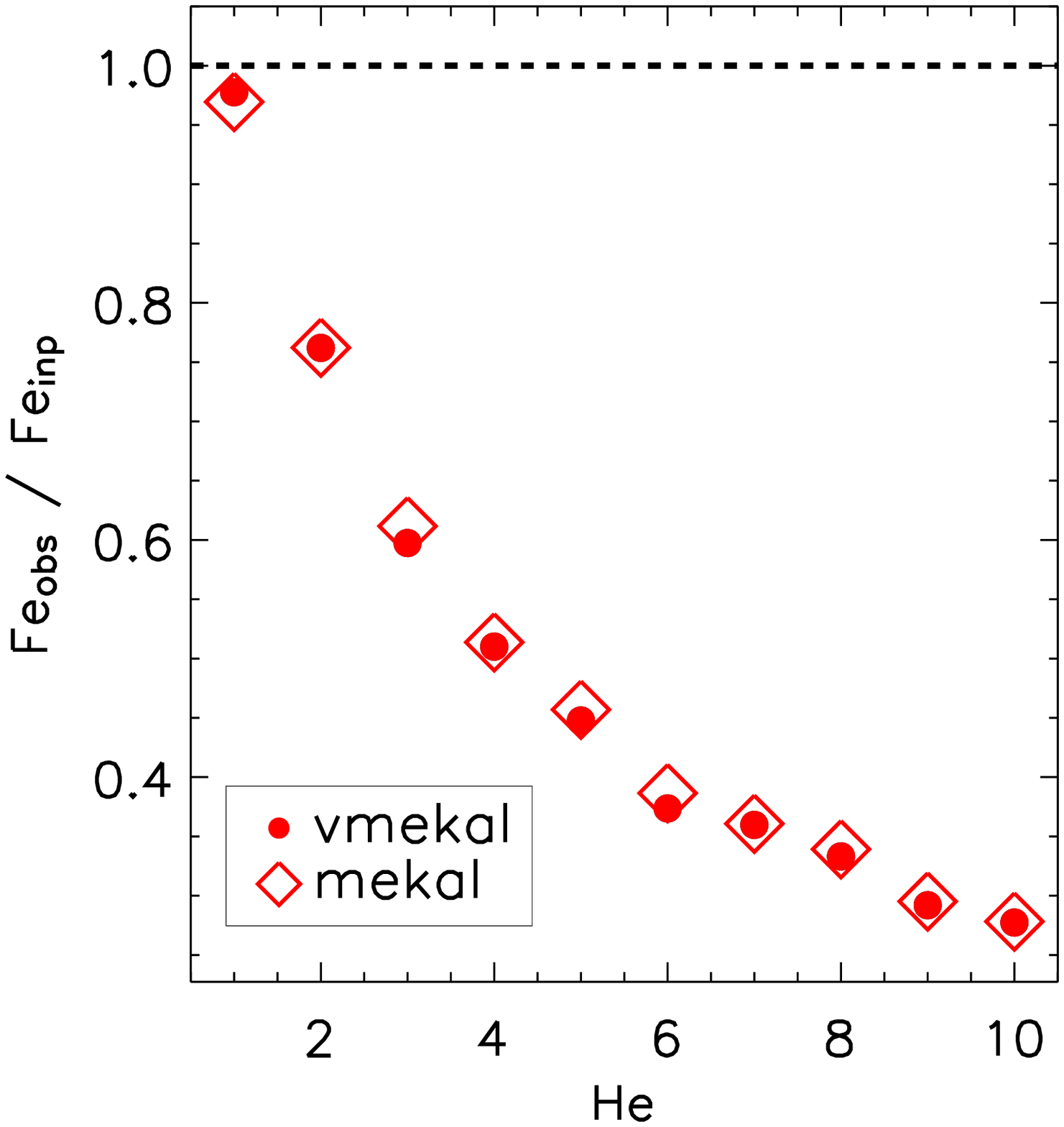,width=.32\textwidth}
\psfig{figure=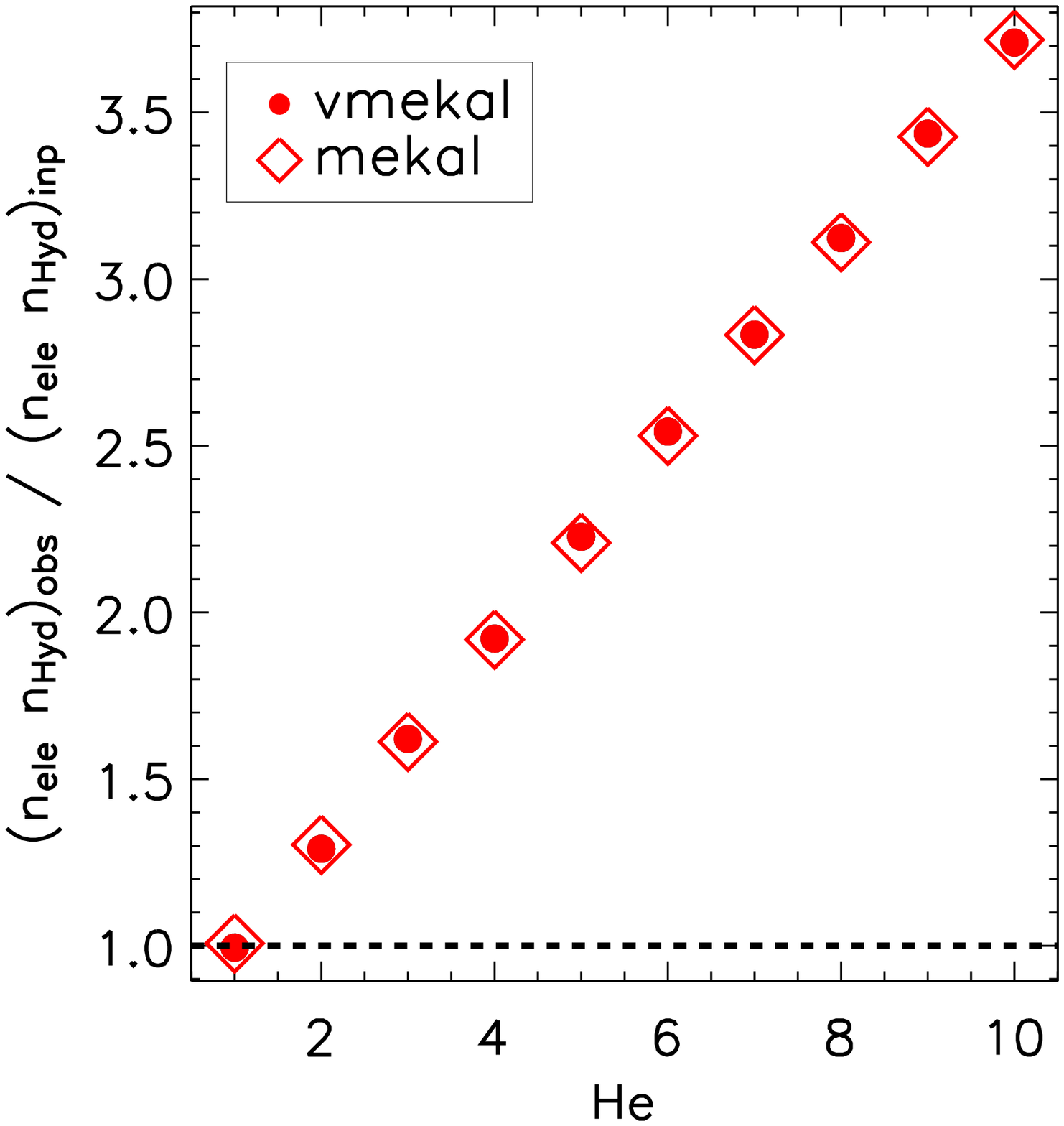,width=.32\textwidth}
\psfig{figure=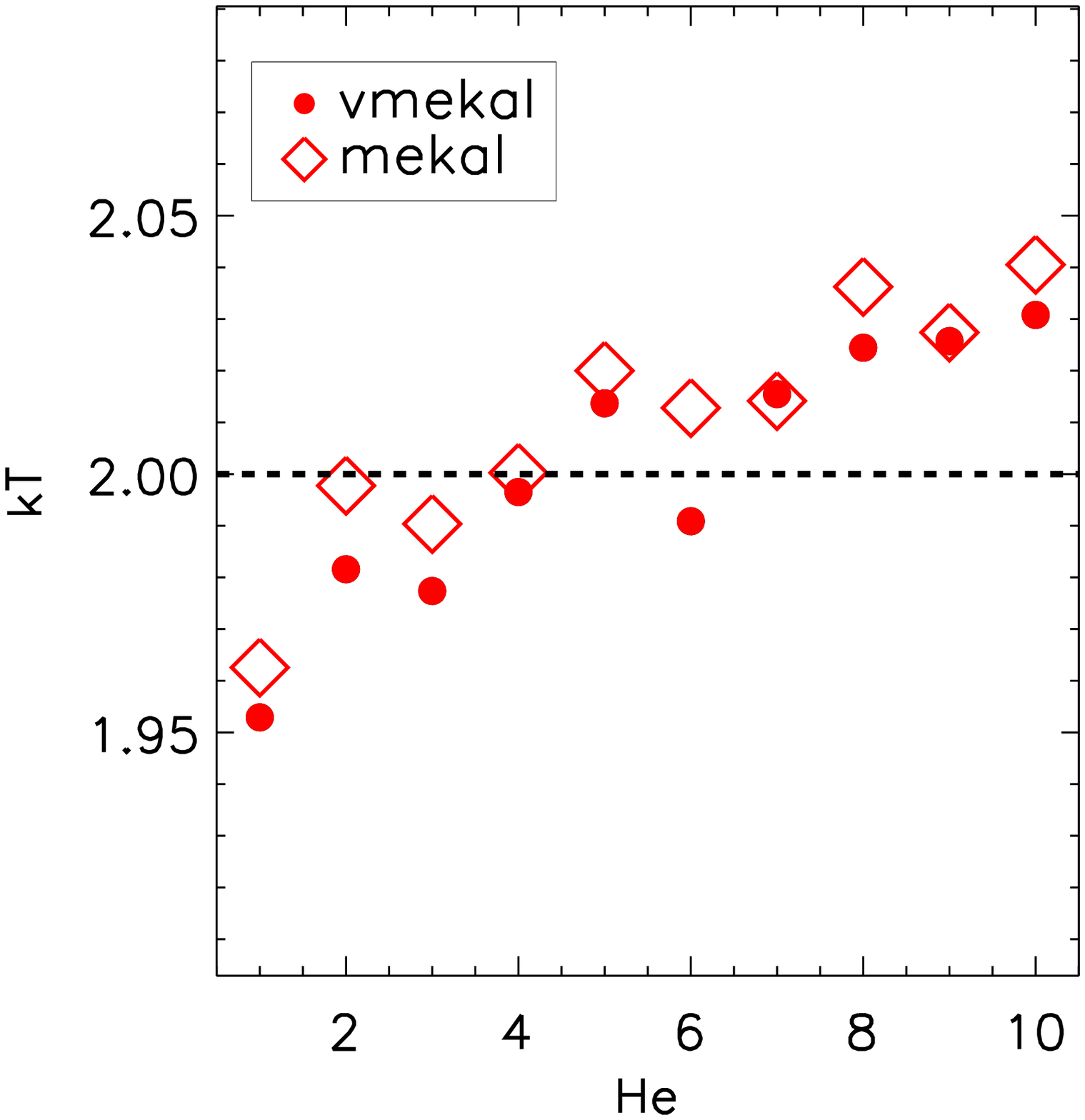,width=.32\textwidth}
}\caption{Best-fit spectral results obtained by fitting {\tt mekal} 
and {\tt vmekal} XSPEC models to simulated {\it Chandra} spectra.
The panels show how the metal (iron) abundance, the 
model normalization and temperature change as function of the helium
abundance considered here between $1$ (the default value) and
$10$ times the solar metal composition as in Anders \& Grevesse (1989).
The results are independent of the input values of the absorbing column 
density, spectra normalization and metal abundance.
} \label{fig:xspec} \end{figure*}

Finally, we have simulated typical thermal spectra of long-time exposed
cluster cores with the ACIS-S configuration on the {\it Chandra} X-ray
satellite generally adopted to fully exploit the larger collecting area
at low energies with respect to ACIS-I.
We have used the {\tt vmekal} model in XSPEC to fake spectra with helium 
abundance in the range 1--10 times the solar abundance as in Anders \& 
Grevesse (1989). Then, we have fitted to the simulated data both {\tt mekal} 
and {\tt vmekal} models with fixed solar abundance of helium, leaving
only the normalization, the temperature and the metal (one single
parameter in {\tt mekal}, 9 parameters --O, Mg, Al, Si, S, Ar, Ca, Fe, Ni--
in {\tt vmekal}) abundance free to vary.
Our results, shown in Fig.~\ref{fig:xspec}, indicate how an unrecognized
super-solar abundance of helium causes underestimates of the metal (iron) 
abundance and overestimates of the model normalization 
owing to the induced increase of the total bremsstrahlung emissivity
(cf. equation~\ref{eq:em}).
We do not measure any relevant variation in the gas temperature and do not
find any significant dependence of these trends upon the assumed values of
column density, normalization and temperature.
An important implication of these results is that it can potentially
explain the drops in metallicity observed through X-ray spectra 
extracted from the inner ($\la 20$ kpc) regions of several 
nearby bright galaxy clusters 
(e.g. Centaurus, Sanders et al. 2002; A2199, Johnstone et al. 2002; 
A3581, Johnstone et al. 2005; A2204, Sanders et al. 2005;
note that also A1795 shows a flat, marginally decreasing metal
abundance in the inner two bins as shown in Ettori et al. 2002; see
few examples of this inverted gradient in Fig.~\ref{fig:abun}).
Figure~\ref{fig:xspec} shows that an enhancement of a factor of 
few ($\sim 2-4$) in the helium abundance with respect to the assumed 
solar value in Anders \& Grevesse (1989)
is sufficient to explain the observed reduction of 20--50
per cent in the metallicity measurements.

\section{Discussion and Conclusions}

We have studied the effects of the sedimentation 
of helium nuclei on the X-ray properties of galaxy clusters.
To this purpose, we have estimated the gravitational acceleration
by adopting functional forms for gas density and temperature profiles
that have been obtained under the assumption that the plasma 
is in the hydrostatic equilibrium with a NFW potential.
These functional forms are fitted to the observed deprojected
gas density and temperature profiles and the best-fit results
are used to model properly in the inner cluster regions the
gravitational acceleration.
The observed gas density and temperature values do not allow
the metal to settle down in cluster cores over timescales
shorter than few $10^9$ years.
On the other hand, by assuming that in the same potential
the gas that is now describing a cool core was initially 
isothermal, the sedimentation times are reduced by 
1--2 order of magnitude within $0.5 r_{\rm s} \approx 0.2 r_{200}$.
The sedimentation of helium can then take place in cluster
cores.
Even modest enhancement in the helium abundance affects 
(i) the relative number of electron and ions $c_{M_i}$, 
(ii) the atomic mean molecular weight $\mu$ and all the 
X-ray quantities that depend on these values, such as
the emissivity $\epsilon$, the gas mass density
$\rho_{\rm g} \propto \mu (1+c_M)$, the total
gravitating mass, $M_{\rm tot} \propto \mu^{-1}$,
the gas mass fraction, $f_{\rm gas} \propto \mu^2 (1+c_M)$.
Moreover, we show that if we model a super-solar abundance 
of helium with the solar value, we underestimate the metal (iron) 
abundance and overestimate the model normalization or emission measure.
For example, with $M_{\rm He} = 3$, the measured iron abundance
and emission measure are $\sim0.65$ and $1.5$ times the input values,
respectively.
Therefore, increased helium abundance might explain the drops
in metallicity observed in the core of some nearby systems.
Other implications of the suspected and plausible super-solar
abundance of helium in the cluster cores are: 
(i) the helium-rich gas will then cool out and form helium-rich stars that
have a typical lifetime shorter and burn hotter than Sun-like objects 
by an order of magnitude (Chris Tout, priv. comm.);
(ii) the central elliptical galaxies of cool core clusters
might be then overabundant in helium
(see speculation on this expected characteristic in Lynden-Bell 1967),
(iii) the central infall of helium induces heating of the surrounding
plasma by $\sim m_{\rm He} n_{\rm He} v_{\rm sed} g \approx 
6.5 \times 10^{-30} M_{\rm He} \left( \frac{n_{\rm H}}{0.01 {\rm cm}^{-3}}
\right) \left( \frac{g}{10^{-8} {\rm erg/g/cm}} \right) 
\left( \frac{v_{\rm sed}}{1 {\rm kpc/Gyr}} \right) {\rm erg/s/cm}^{3}$ 
that can offset the central cooling of about 
$n_{\rm H} n_{\rm e} \Lambda(T) \approx 2 \times 10^{-27} 
\left( \frac{n_{\rm H}}{0.01 {\rm cm}^{-3}} \right)^2 {\rm erg/s/cm}^{3}$ 
by a negligible amount. 
On the other hand, the diffusion of helium can be suppressed by
the action of confinement due to reasonable magnetic fields,
as recently pointed out from Chuzhoy \& Loeb (2004),
or limited to the very central region ($r<20$ kpc)
by the turbulent motion of the plasma.

\vspace*{-1truecm}
\section*{ACKNOWLEDGEMENTS} ACF acknowledges the support
of the Royal Society.
We thank Jeremy Sanders and Roderick Johnstone 
for providing computer-readable tables of the gas
density, temperature and metallicity profiles of Centaurus and A2199.
We thank the referee, Eugene Churazov, for insightful comments.

\end{document}